\documentclass[preprint]{aastex}
\shorttitle{Stability of Standing Accretion Shocks}
\shortauthors{Blondin, Mezzacappa, and DeMarino}
\slugcomment{To appear in The Astrophysical Journal}

\begin{document}        
\title{STABILITY OF STANDING ACCRETION SHOCKS, WITH AN EYE TOWARD
CORE COLLAPSE SUPERNOVAE}
\author{John M.~Blondin}
\affil{Department of Physics, North Carolina State University, Raleigh, NC 27695-8202}
\email{John\_Blondin@ncsu.edu}
\author{Anthony Mezzacappa}
\affil{Physics Division, Oak Ridge National Laboratory, Oak Ridge, TN 37831-6354}
\email{mezzacappaa@ornl.gov}
\and
\author{Christine DeMarino}
\affil{Department of Physics, North Carolina State University, Raleigh, NC 27695-8202}

\begin{abstract}
We examine the stability of standing, spherical accretion shocks.
Accretion shocks 
arise in core collapse supernovae (the focus of this paper), star formation, 
and accreting white dwarfs and neutron stars.
We present a simple analytic model and use time-dependent hydrodynamics
simulations to show that this solution
is stable to radial perturbations.  In two dimensions 
we show that small perturbations
to a spherical shock front can lead to rapid growth of turbulence
behind the shock, driven by the injection of vorticity from the 
now non-spherical shock. We discuss the ramifications this instability
may have for the supernova mechanism.

\end{abstract}        
\keywords{accretion---hydrodynamics---shock waves ---supernovae:general---turbulence}
       
\section{INTRODUCTION}

Accretion shocks occur in many settings in astrophysics, including
core collapse supernovae, star formation, and accreting compact 
objects. In many cases these accretion shocks may exist in a 
quasi-steady state, with the postshock
flow gradually settling onto the accreting object such that the 
radius of the accretion shock remains constant: a standing accretion shock,
or SAS.

A particularly interesting example of a standing accretion shock arises
in core-collapse supernovae.  In classical models of such supernovae, 
an expanding shock front stalls at a radius $\sim 100-200$ km, and
remains relatively stationary for an eternity $\sim 300$ ms
\citep[see, for example,][and references therein]{mez01}, 
during which time it is ``revived'' by an as yet unknown combination
of factors including neutrino heating, convection, rotation, and
magnetic fields.

Despite the ubiquity of standing accretion shocks in astrophysical lore,
there is little discussion in the literature of their stability.  
This oversight may be attributed to the fact that strong,
adiabatic shocks are known to be self-healing in that small perturbations
to a planar shock front lead to postshock flow that acts against the
perturbations \citep[e.g.,][]{w74}.  The situation may be different in the case 
of a spherical accretion shock.  The convergence of the postshock flow
may amplify perturbations, and the 
confined postshock volume increases the opportunity
for the postshock flow to feed back on the dynamics of the shock front.
For example, an analysis of a converging blastwave
suggests that such shocks are unstable for high Mach numbers \citep[e.g.,][]{w74},
in contrast with planar shocks.
More recently, 
\citet{fog02} has described a vortical-acoustic instability of spherical
accretion shocks in the context of accreting black holes.  In this case
the instability results from a feedback between acoustic waves perturbing
the spherical shock and the perturbed shock exciting vorticity perturbations
that advect downstream towards the accreting object.  These
vorticity waves generate acoustic waves that propagate back to the 
shock, completing the feedback cycle.  

In this paper we investigate the stability of an SAS and, in particular,
the role of oblique (aspherical) shocks in feeding turbulence in the
confined postshock region.  To this end we present one- and
two-dimensional time-dependent hydrodynamics simulations of an idealized
SAS.  We begin by defining an idealized SAS and presenting an analytic
solution for the postshock flow in Section 2.  In Section 3 we describe
the numerical method we use to evolve a time-dependent hydrodynamics
simulation of an SAS, and in Section 4 we use these simulations to
demonstrate the stability of an SAS in one dimension.  In Section 5 we
describe the unstable nature of an SAS in two dimensions.  We conclude
by discussing the relevance of an unstable accretion shock to the
problem of core-collapse supernovae.

\section{IDEALIZED MODEL OF A STANDING ACCRETION SHOCK}

To model standing accretion shocks, we consider an idealized adiabatic
gas in one dimension accreting onto a star of mass $M$.
We begin by assuming that the infalling gas has had time to accelerate
to free fall and that this velocity is highly supersonic.  The fluid
variables (velocity, $u$; density, $\rho$; pressure, $p$) 
just behind the standing accretion shock are then given by
the Rankine-Hugoniot equations:
\begin{equation}
u = -\frac{\gamma - 1}{\gamma + 1}(2GM)^{1/2}R_s^{-1/2}
\end{equation}
\begin{equation}
\rho = \frac{\gamma + 1}{\gamma - 1}
\left(\frac{\dot M}{4\pi}\right)(2GM)^{-1/2}R_s^{-3/2}
\end{equation}
\begin{equation}
p = \frac{2}{\gamma + 1}
\left(\frac{\dot M}{4\pi}\right)(2GM)^{1/2}R_s^{-5/2}
\end{equation}

\noindent where $\gamma$ is the adiabatic index and $R_{s}$ is the
shock radius.

Below the shock front we assume radiative losses are negligible (we
will discuss the appropriateness of this assumption, particularly in 
the case of core collapse supernovae, in more detail later) and
the gas is isentropic, such that $p/\rho^\gamma$ is constant.  
The assumption of a steady-state isentropic flow implies there is 
a zero gradient in the entropy of the postshock
gas, which in turn implies this flow is marginally stable to 
convection.  This allows us to separate effects due to 
convection from other aspects of the multidimensional fluid 
flow. 
In the more general case where a negative entropy gradient might
drive thermal convection \citep{hbhfc94,bhf95,jm96,mcbbgsu98b,fh00},
such convection could act as a seed for
the instabilities we will examine in this paper.

Under these assumptions, the postshock flow is described by
the Bernoulli equation:
\begin{equation}
\frac{1}{2}u^2 + \frac{\gamma}{\gamma-1}\frac{p}{\rho}-\frac{GM}{r}=0,
\end{equation}
where the constant on the right hand side is set by the assumption of
highly supersonic freefall above the accretion shock 
($P/\rho \ll \frac{1}{2} u^2 \approx GM/r$).

We normalize the problem such that $GM=0.5$, $\dot M=4\pi$, and the
accretion shock is at
$R_s=1$.  Rewriting the Bernoulli
equation in terms of velocity and radius, we have
\begin{equation}
r u^2 + \alpha r^{3-2\gamma} u^{1-\gamma} - 1 = 0,
\label{eq:vel}
\end{equation}
where
\begin{equation}
\alpha = \frac{4\gamma}{(\gamma+1)(\gamma-1)}\left(\frac{\gamma-1}{\gamma+1}
\right)^{\gamma}.
\end{equation}
For $\gamma = 5/3$, one can write down an analytic expression for
the post-shock gas:
\begin{equation}
u(r) =  \frac{\gamma-1}{\gamma+1} r^{-1/2}.
\end{equation}
In this special case, the Mach number of the postshock flow remains
fixed at $1/\sqrt 5$, as the gas accelerates inwards at a fixed fraction
of the freefall velocity.  

\begin{figure}[!hbtp]
\epsscale{0.70}
\plotone{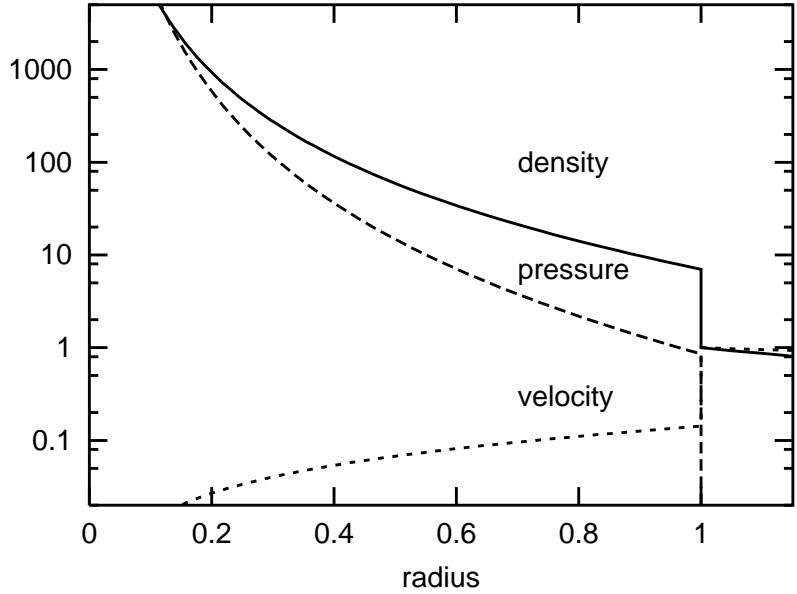}
\caption{Analytic solution for an adiabatic standing accretion shock
assuming an adiabatic index $\gamma = 4/3$.
The density, pressure, and velocity (absolute value) are integrated inwards from the
accretion shock at $R_s=1$.}
\label{fig:analytic}
\end{figure}

The behavior of the postshock gas 
is markedly different for smaller values of $\gamma$.
Below a critical value, 
$\gamma_c = 1.522$, the gas decelerates, rather than accelerates, 
as it drops below
the standing accretion shock.  
This isentropic settling solution 
below the accretion shock is shown
in Figure \ref{fig:analytic} for an adiabatic index $\gamma = 4/3$.

\begin{figure}[!hbtp]
\epsscale{0.70}
\plotone{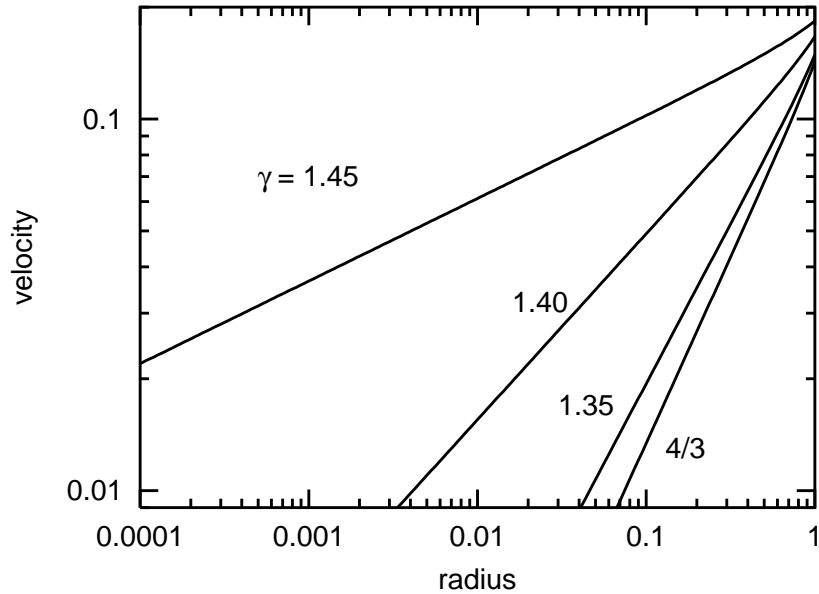}
\caption{The infall velocity of the settling solution for different
values of the adiabatic index, $\gamma$.  At radii much smaller
than the shock radius, the solutions follow a power-law in radius.}
\label{fig:1dsolutions}
\end{figure}

The behavior of these settling solutions at 
small radii can be found by neglecting the first term in eqn. \ref{eq:vel},
which corresponds to the kinetic energy of the gas.  This amounts to
assuming that 
the local Mach number has become very small.  The resulting solution
represents an approximately hydrostatic atmosphere with a constant
mass inflow rate.  The velocity is then a power-law in radius:
\begin{equation}
u(r) \propto r^ \frac{3-2\gamma}{\gamma-1}.
\end{equation}
For large values of $\gamma$ (but still below the critical
value of 1.522) this is a relatively shallow power law,
and the infalling gas slows down only when very close to $r=0$.  For
soft equations of state, however, the gas decelerates relatively quickly.
To maintain the same pressure gradient required for (nearly) hydrostatic
equilibrium, a gas with a softer equation of state must be compressed 
faster.  This translates into a faster deceleration under the assumption
of a constant mass inflow rate.
For $\gamma = 4/3$, the infall velocity drops linearly with radius; at
$r=0.1$, the gas is settling inwards with a velocity of only 1\% the
local sound speed.  These settling solutions are shown in 
Figure \ref{fig:1dsolutions} for various values of $\gamma$.
We note that these solutions are the same as those used by
\citet{chev89} to model the post-supernova fall-back of stellar
gas onto the neutron star.

\begin{figure}[!hbtp]
\epsscale{0.70}
\plotone{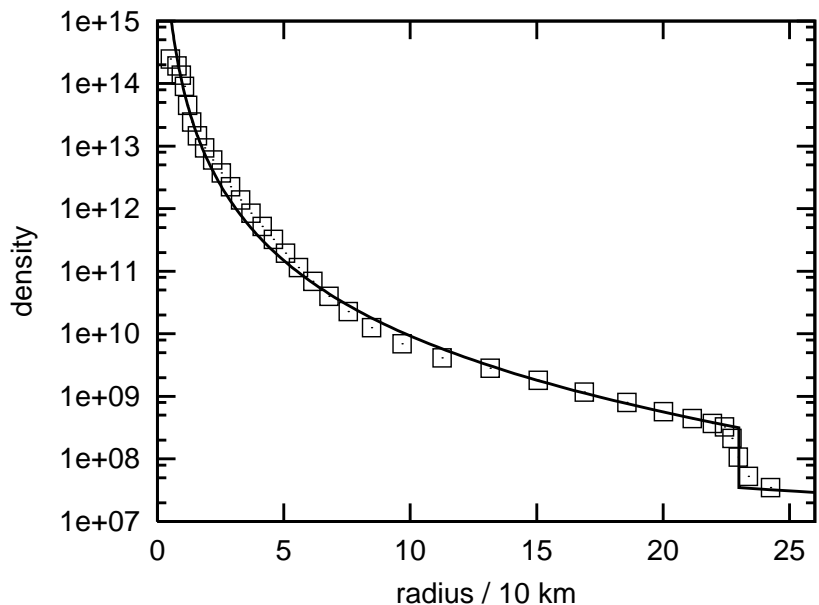}
\caption{The idealized SAS solution with $\gamma=1.25$ (solid line)
provides a reasonably good approximation to a more realistic
supernova model. }
\label{fig:snfit}
\end{figure}

For comparison, in Figure \ref{fig:snfit}
we plot our model SAS solution along with
the results of a complete one-dimensional model of stellar core 
collapse, bounce, and the postbounce evolution,
which includes Boltzmann neutrino transport and a realistic equation of
state.
These results show the structure of the stalled supernova shock
at roughly 300 ms after core bounce, as described in \citet{mez01}.
Our SAS solution has been
scaled to match the postshock density of the supernova model.  For this
particular scaling, the unit of time in our normalized SAS model 
corresponds to 6 ms in the supernova model.

\section{NUMERICAL MODEL}

We use the time-dependent hydrodynamics code VH-1 
(\url{http://wonka.physics.ncsu.edu/pub/VH-1/}) to study the 
dynamics of a standing accretion shock.  For this problem we have
modified the energy conservation equation in VH-1 to evolve a total energy
that includes the time-independent gravitational potential:
\begin{equation}
E_{tot} = \frac{1}{2}u^2 + \frac{1}{\gamma -1}\frac{p}{\rho} + \Phi
\end{equation}
where $\Phi = -GM/r$ for the assumed gravitational field of a point mass.
This approach ensures conservation of the total energy to within 
machine round-off error.

The numerical simulations are initialized by mapping the analytic
solution onto a grid of 300 radial zones extending from
$0.1 R_s$ to $2 R_s$.  The radial width of the zones increases
linearly with radius such that the
local zone shape is roughly square in the two-dimensional simulations; 
i.e., $\Delta R \approx R\Delta\theta$,
to minimize truncation error.  The two dimensional simulations 
used 300 angular zones covering a polar
angle from $0$ to $\pi$, assuming axisymmetry about the polar axis.
The fluid variables
at the outer boundary are held fixed at values appropriate for
highly supersonic free-fall at a constant mass accretion rate,
consistent with the analytic standing accretion shock model.

For the purpose of our numerical simulations we must impose an inner
boundary, which one would nominally associate with the surface of the
accreting object.  This boundary must allow a mass and energy flux off
the numerical grid in order to maintain a steady state.  We do this by
imposing a leaky boundary condition: an inner boundary with a specified
radial inflow velocity corresponding to the chosen radius of the inner
boundary.

Our numerical code employs a Lagrange-remap algorithm, in which the
Lagrangian fluid equations are evolved using solutions to the Riemann
problem at zone boundaries.  We fix the velocity at the inner boundary
by replacing the Riemann solution for the time-averaged velocity at the
inner boundary, with a fixed value.  This value of the boundary flow
velocity is chosen based on the analytic value at the boundary radius (see
Figure 2), and, in effect, this value sets the height of the standing
shock.  The pressure gradient at the inner boundary is set to match the
local gravitational acceleration, mimicking a hydrostatic atmosphere
(the SAS is nearly hydrostatic far below the shock).  The density at
this leaky boundary is adjusted to force a zero gradient, allowing a
variable mass flux off the inner edge of the numerical grid.  This
specific set of boundary conditions was influenced by the desire to
construct time-dependent solutions that are stable in one dimension
(see Section 4).

To test the dependence of our results on this inner radial boundary
condition we also ran simulations in which we replaced the leaky
boundary with a hard reflecting surface and added a cooling layer
above it.  This cooling layer is meant to mimic the energy loss due to
neutrino cooling above the surface of the proto-neutron star in a
more realistic model.  Stationary accretion shock solutions with a
cooling layer are described in \cite{hc92}. We have chosen a cooling
function described by $\alpha=\beta=2.5$ in their notation.  These values
correspond to a radiation-dominated gas ($p\propto T^4$) loosing energy
through neutrino cooling with a negilible density dependence but 
a strong temperature dependence 
($\dot E \propto T^{10}$).   We adjusted 
the magnitude of the cooling function to place the standing shock at
a radius of unity.  

For the two-dimensional simulations,
reflecting boundary conditions were applied at the polar boundaries
and, the radial boundaries were treated the same as in the one-dimensional
simulations.  The tangential
velocity was initialized to zero everywhere in the computational 
domain.  The tangential velocity at the inner radial boundary was
set to zero throughout the simulations.

The PPM algorithm \citep{cw84} used in VH-1 is a good method to study the 
stability
of accretion shocks because shock transitions are typically confined
to only two numerical zones.
However, a drawback of this algorithm is the generation of postshock
noise when a strong shock aligned with the numerical grid moves 
very slowly across the grid.  This is exactly the situation encountered
in this problem.  In one dimensional simulations this produces small
entropy waves that advect downstream from the shock with little effect
on the flow.  In multiple dimensions, the zone-to-zone fluctuations in
pressure and density transverse to the shock front generate
considerable postshock noise \citep{lev98}.

To minimize the effects of this numerical noise we have added a small
amount of dissipation by ``wiggling'' the computational grid in both
the radial and azimuthal directions \citep{cw84}.  The radial grid
``wiggling'' is confined to zones near the location of the steady-state
shock so that our inflow and outflow boundaries are not affected.  This
approach reduces the entropy variations in one-dimensional flows by an
order of magnitude, and in two dimensions the flow remains stable
(nearly spherical) for many sound crossing times (defined here as 
the time for a sound wave to cross the radius of the accretion shock).  
In the two-dimensional
case, the smooth flow is eventually disturbed by noise generated at the
corners of the numerical grid, where the polar axes meet the inner
radial boundary.  This numerical noise ultimately seeds the feedback
mechanism discussed in Section 5, and the flow becomes unstable.

\section{STABLE FLOW IN ONE DIMENSION}

Before jumping to the effects of aspherical accretion shocks, we first
examine the stability of our idealized model in one dimension.  Our
goals here are two-fold: (1) to test the ability of our numerical model
(in particular, our inner boundary conditions) to simulate a standing
accretion shock and (2) to determine the stability properties of our
solutions under radial perturbations.

If the initial conditions are set to match the inner boundary 
conditions (i.e., the initialized flow velocity at the inner 
boundary equals the imposed value enforced by the boundary conditions), 
our simulations are relatively uneventful.  After
initial transients die away, the
solutions settle into a steady-state that closely matches the
analytic solution.

\begin{figure}[!hbtp]
\plotone{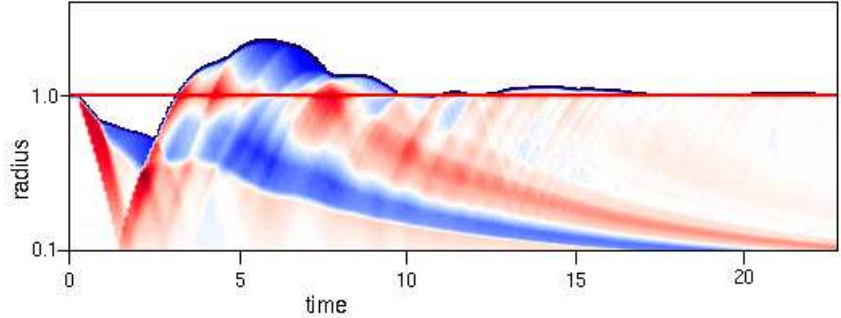}
\caption{The pressure deviations from the analytic settling solution
for a simulation in which a shell with 3 times the ambient density 
is dropped through the accretion shock.  Red (blue) represents pressures higher
(lower) than steady-state.  The solid horizontal line marks the location
of the standing accretion shock at $R=1$. The thin blue curve along
the top 
is the radial trajectory of the perturbed shock.
Note the strong reflection of pressure waves at 
the inner boundary at $t\sim 1.5$.
The unit of time in our normalized SAS model 
corresponds to $\sim 6$ ms in the supernova model shown in Figure 3.}
\label{fig:1dpert}
\end{figure}

To test the stability of these models, we tried various ways
of perturbing the initial solution.  In one set of experiments we created
a mismatch between the initial conditions and the imposed flow
velocity at the inner boundary.  If the fixed boundary velocity is
set smaller than the flow velocity in the initial analytic solution,
the accretion flow is choked off at the inner boundary, and the buildup
of pressure initially drives the accretion shock to larger radii.
In another set of experiments, we added a large density perturbation
in the supersonic free-fall portion of the initial conditions.  When 
this over-dense shell reaches the accretion shock, the increased ram pressure
initially drives the shock down to smaller radii.

The results of one of these attempts to perturb the SAS are shown in Figure \ref{fig:1dpert}.
In all cases the shock front oscillated a few times, but these oscillations
were always strongly damped.  For the $\gamma=4/3$ solution shown here, 
the time for material to drift from
the accretion shock to the inner boundary is $\tau_d = 17$, compared to
a sound crossing time of $\tau_s =  1.6$ and a freefall time of $\tau_f = 0.65$. 
The shock front rebounds from a perturbation on the relatively short
sound-crossing timescale, $\tau_s$.  For large perturbations 
this rebound typically overshoots the equilibrium shock radius and 
subsequent oscillations decay away 
after several $\tau_s$.  The accretion shock radius
always settled back to within $0.5\%$ of the original steady-state radius
after several $\tau_s$.
The mass accretion rate takes longer to settle down to the steady-state value,
as the perturbed postshock flow gradually advects down through the leaky 
boundary.  It is only after a few $\tau_d$ that
the full effects of the perturbation have
advected out of the region and the mass accretion rate
at the inner boundary has relaxed to the steady-state value.

These simulations demonstrate that SAS models are stable to radial
perturbations.  As long as the flow remains spherically symmetric, the
accretion shock remains at a fixed radius.

\section{INSTABILITY IN TWO DIMENSIONS}
 
To study the stability of standing accretion shocks under the
assumption of axisymmetry we ran additional simulations in two dimensions
with a variety of initial perturbations,
including small density perturbations in the
infalling material, weak pressure waves emanating from near the
inner boundary, and small random velocity fluctuations in the postshock flow.  
To quantify the behavior of these accretion simulations our
numerical code kept track of several global metrics, including
the total kinetic energy (and the component in the angular direction),
the thermal and gravitational potential energy of the shocked gas, 
the total vorticity, 
the average radius of the accretion shock, and the angular variance in 
the accretion shock radius. 

\begin{figure}[!hbtp]
\epsscale{0.70}
\plotone{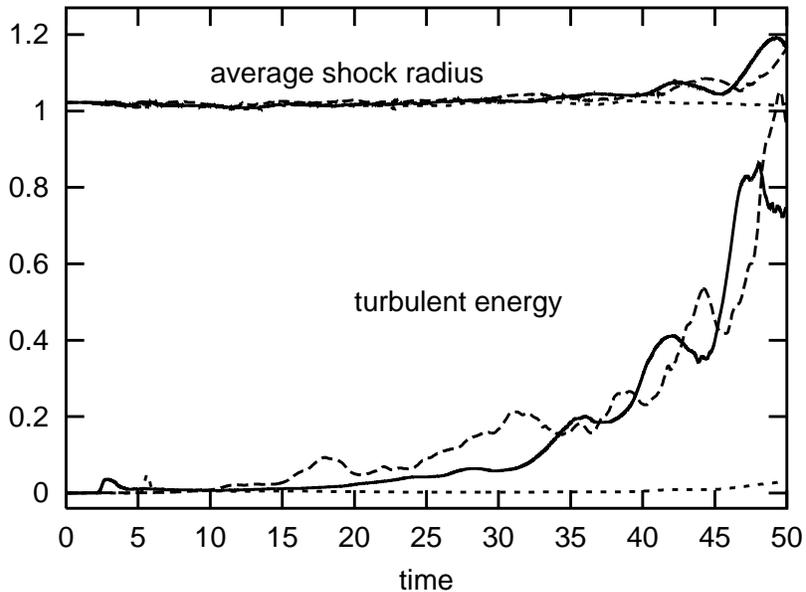}
\caption{The growth of interior turbulent energy (defined here as
the total kinetic energy in the angular direction) is accompanied by
an increase in the average shock radius.  The two short-dashed lines
correspond to the two-dimensional 
simulation with spherically symmetric infall, which remains nearly
symmetric throughout the evolution.  The
two solid lines correspond to a two-dimensional 
simulation initialized with a small asymmetric perturbation.  
The two long-dashed lines correspond to a 
similar perturbed model except that the leaky inner boundary
has been replaced with a hard reflecting boundary and a thin 
cooling layer.}
\label{fig:turb}
\end{figure}

\begin{figure}[!hbtp]
\plotone{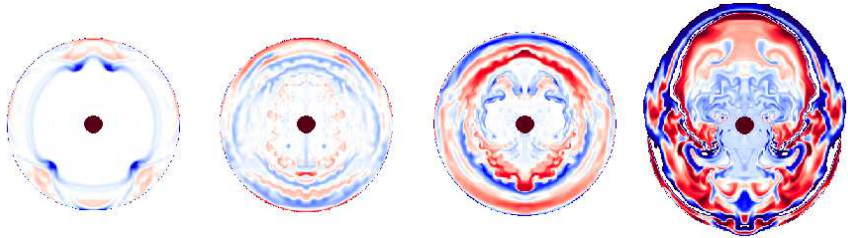}
\caption{Images of the gas entropy (red is higher than the equilibrium
value, blue is lower) illustrate the instability of
a spherical standing accretion shock. 
This model has $\gamma=4/3$ and is perturbed by placing over-dense 
rings into the infalling preshock gas.   Note that with the 
scaling for
a realistic supernova model, the last image on the right corresponds 
to $\sim 300$ ms.
These simulations are
axisymmetric, forcing a reflection symmetry about the vertical axis.}
\label{fig:k1a01}
\end{figure}

All of our two-dimensional SAS simulations proved to be unstable.
To illustrate the typical behavior of an axisymmetric accretion shock
we will focus on a model with $\gamma=4/3$ and initialized
with two rings (a density enhancement of 20\%) placed asymmetrically 
in the preshock flow (see Figure \ref{fig:k1a01}).  
For sufficiently small perturbations, 
the early evolution is characterized by sound waves bouncing throughout
the interior of the accretion shock.  These waves are nearly spherically
symmetric, but gradually take on an asymmetry characterized by 
a spherical harmonic with $l=1$.
The strength of these asymmetric waves
grows with time, and eventually they begin to significantly impact
the global properties of the accretion shock.  This evolution is
shown in Figure \ref{fig:turb}, where we plot the total energy
in angular motion (which we assume to be proportional to the local
turbulent energy) and the angle-averaged shock radius.  
We also show the results for an unperturbed simulation to show
that our numerical techniques can hold a SAS stable for at least 
several dynamical times.  From this, we conclude that the introduction
of asymmetric perturbations leads to a growing postshock turbulence
in standing accretion shocks.

Also shown in Figure \ref{fig:turb} is a similar 
simulation but with different inner boundary conditions: a reflecting
wall plus a thin cooling layer.  The initial radial solution is
nearly identical to the adiabatic solution except close to the
inner boundary, where the cooling becomes strong and the gas quickly
settles into a thin layer on top of the hard surface.  As shown  
in  Figure \ref{fig:turb}, the evolution is qualitatively the same
as in the adiabatic model with a leaky boundary.  This comparison verifies
the independence
of these results from the details of the boundary conditions.

To further illustrate this early evolution we show a time series of
the two-dimensional flow in Figure \ref{fig:k1a01}.
We use the gas entropy to visualize the flow because a steady, spherical
shock would produce a postshock region of constant entropy.  Thus, any
deviation from a spherical shock shows up as a change in the postshock
entropy.  Aside from the overall trend of growing perturbations in the
postshock entropy, this figure illustrates the dominance of low-order modes.
This is a generic result of our two-dimensional simulations.  While the
wave modes in the early evolution depend on the initial perturbations, the 
$l$ = 1 mode always grows to dominate the dynamics.

\begin{figure}[!hbtp]
\plotone{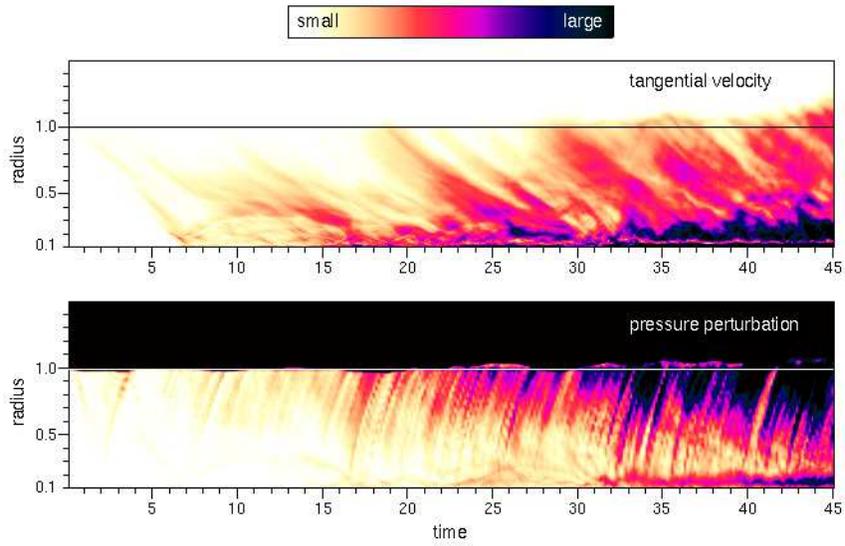}
\caption{The time evolution of a 2D simulation is illustrated in
spacetime diagrams.  The color at a given radius and time corresponds
to the angle-average of the absolute value of the
tangential velocity (top) and absolute value of the
deviation of pressure from the steady-state
solution, $\Delta P/P$ (bottom).  Regions 
that are under-pressured with respect to the steady-state solution 
collect near the bottom of the ``atmosphere'' after $t\sim 15$ 
and correspond to the 
centers of strong vorticies that have become trapped at these radii
by advection.  Direct evidence of these vortices are seen in the 
velocity plot.}
\label{fig:spacetime}
\end{figure}

To better understand the origin of the SAS instability we show a spacetime
diagram of angle-averaged quantities in Figure \ref{fig:spacetime}.
Here we see two dominant trends: (1) pressure waves rising up from
small radii at the local speed of sound to
perturb the shock front and (2) regions of high transverse velocity
originating at the shock and advecting downwards at the local flow
speed (slower than the sound speed).  It is the feedback
between these two waves (at least for low values of $l$) 
that leads to amplification of the initial
perturbations.  Aspherical pressure waves rising up to the 
accretion shock distort the shock front from its equilibrium spherical 
shape.  The slight obliquity of the 
shock front relative to the radial pre-shock flow
then creates non-radial post-shock flow that advects inward, feeding
turbulent flow at small radii.  This growing turbulence in turn
drives stronger aspherical pressure waves that propagate upwards and
further distort the accretion shock.  This feedback was discovered
independently in the context of accretion onto a black hole by 
\citet{fog02}, who dubbed it a ``vortical-acoustic'' cycle. 

To contrast this with planar flow, any vorticity generated by perturbations 
to the planar shock front are advected downstream, unable to impact
the shock dynamics.  In the case of an SAS, however, the vorticity is
trapped in a finite volume.  With nowhere to go, the vorticity
grows with time and eventually becomes strong enough to affect
the shock front.

\begin{figure}[!hbtp]
\epsscale{0.70}
\plotone{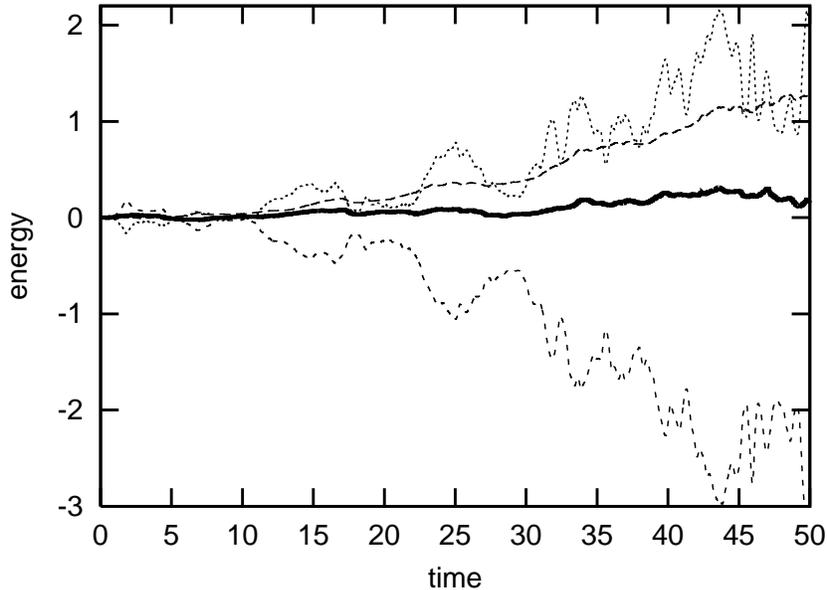}
\caption{The change in energy ($E - E_o$) of the postshock gas is shown
for the total (solid), kinetic (long dash), thermal (short dash), and 
gravitational potential (dotted)
energies.  The initial values, $E_o$, are -15.8, +0.08, +47.6, and -63.5,
respectively.}
\label{fig:energy}
\end{figure}

\begin{figure}[!hbtp]
\epsscale{0.70}
\plotone{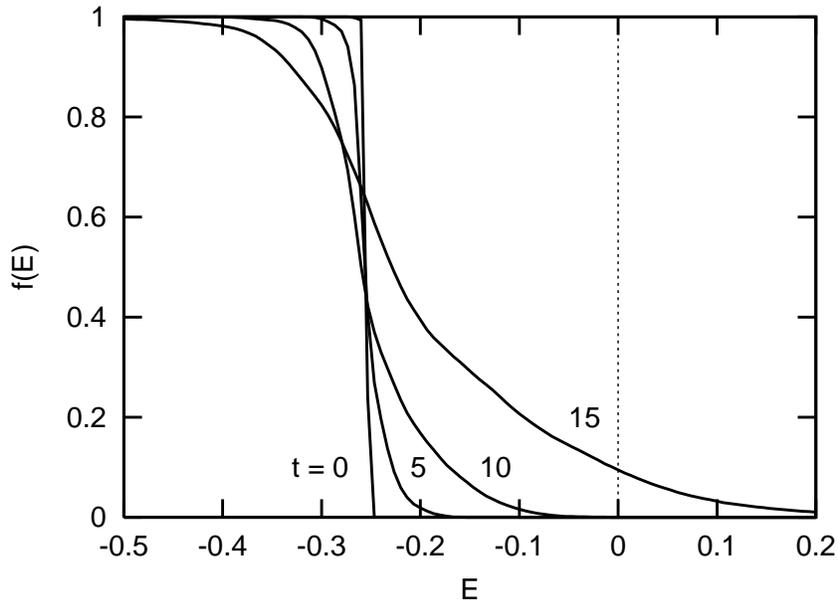}
\caption{We plot 
the mass fraction of postshock gas with total energy per unit mass 
(scaled to the local gravitational potential energy) above a
given value, $E$.  In the initial SAS solution, all of the postshock
gas has a total scaled energy per unit mass of $E=-0.245$ for $\gamma = 4/3$; 
therefore, the sequence shown in this figure 
begins as a step function at that value.  With time, the gas becomes
redistributed in energy, with some gas having more negative total energy 
and some gas having substantially more positive total energy.
At late times, a significant fraction of the postshock gas has $E>0$.
}
\label{fig:energyfrac}
\end{figure}

Once the turbulent energy density in the postshock flow becomes
significant, the volume of the shocked gas begins to 
grow with no apparent bound.  
The origin of this increase in shock radius can be traced
to the distribution (kinetic versus thermal versus gravitational) 
of total energy below the shock.  Note that
this is strictly adiabatic flow---there is no source of external
heating to assist in driving the shock outwards.
The total energy on the computational grid remains relatively
constant (the total energy of the gas flowing off the inner edge
of the grid is small and varies slightly due to the growing turbulence, 
while the total energy of the gas
falling onto the grid from the outer boundary is approximately zero),
but the distribution of energy changes.  As the nominally spherical
accretion shock becomes more distorted, the postshock gas retains
more kinetic energy at the expense of thermal energy.  This gradual
change can be seen in a plot of the various components of the
total energy (the three terms in eqn. 9) integrated over
the volume of shocked gas. This is shown in Figure \ref{fig:energy}. 
Moreover, this redistribution of energy among
kinetic, thermal, and gravitational components results in a
redistribution of the total energy per unit mass in the postshock gas.
In Figure \ref{fig:energyfrac} we see that, as the instability 
described here develops, an increasing fraction of the postshock gas 
attains positive energies at the expense of gas that becomes more 
bound.  There is no new source of energy in this system, only a 
redistribution of energy.

\begin{figure}[!hbtp]
\plotone{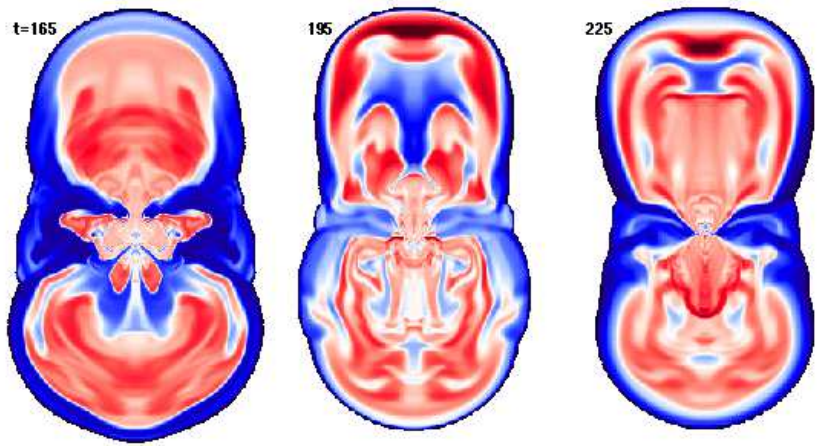}
\caption{The self-similar growth of the asymmetric accretion
shock is illustrated in these entropy images from three different
times in our simulation.  The postshock volume grows with 
time, but each successive image is scaled 
down by a factor of two to produce three images of 
roughly the same size.  Here blue (red) represents low (high) entropy.}
\label{fig:selfsim}
\end{figure}

To follow the long-term evolution of an SAS we continued the simulation
of our standard model, with $\gamma = 4/3$, but allowed the numerical
grid to expand radially to accommodate the increasing shock radius.
We followed the expansion out to a radius of 15 in this fashion.  
The evolution can be broken into three distinct epochs.  At early
times ($t<40$) the shock remains roughly spherical, and the volume
of shocked gas is approximately constant while the interior turbulent
energy grows steadily.  The intermediate epoch is characterized by
large oscillations in the postshock flow that lead to an aspect ratio
for the accretion shock that varies between one and two.  The volume
of shocked gas exhibits a growing oscillation, reminiscent of a 
breathing mode.  This phase of the evolution is dominated by the 
$l$ = 1 mode; i.e., is distinctly oscillatory.
Beyond a time of 170, the flow takes on a strikingly self-similar
form dominated by the $l$ = 2 mode; i.e., is distinctly bipolar.  
The aspect ratio becomes remarkably constant at a value 
$\sim 2$ (2.3 to be precise), but the scale of this aspherical shock 
grows roughly
linearly with time.  From $t=165$ to $t=225$, the angle-averaged
shock radius increases from 5 to 15.  This corresponds to an
expansion velocity of order 50\% the escape velocity at the shock.
This self-similar expansion is illustrated in Figure \ref{fig:selfsim},
where we plot images of the gas entropy at three different times
chosen such that the shock has roughly doubled in size in each 
interval.
When scaled to the same size, the images look qualitatively similar.

\begin{figure}[!hbtp]
\epsscale{0.70}
\plotone{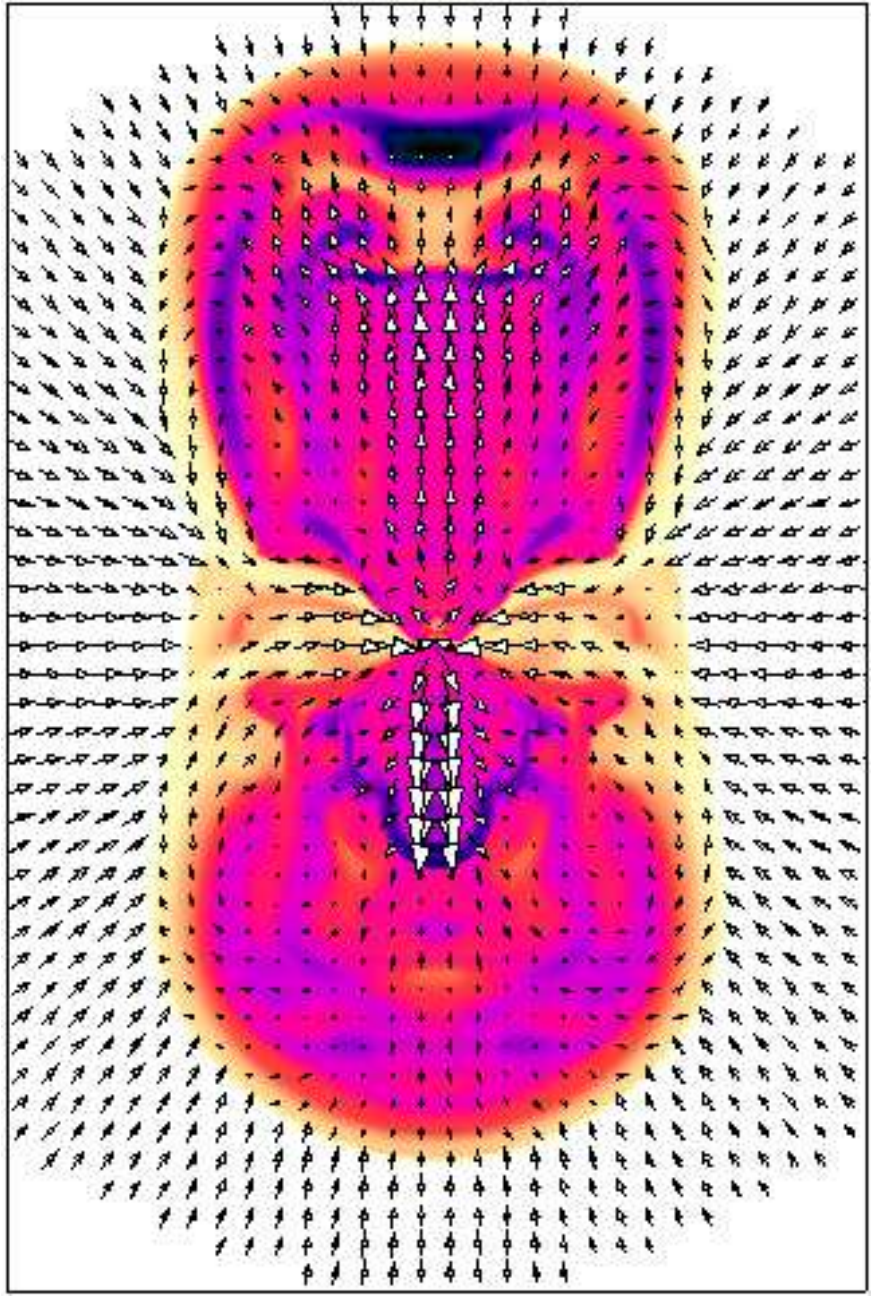}
\caption{The expansion of the accretion shock is driven by
accretion in the equatorial plane feeding quasi-periodic 
outflows along 
the symmetry axis.  The color scale in this image depicts
gas entropy, with dark colors representing high entropy.}
\label{fig:latetime}
\end{figure}

The dynamics of this self-similar state are illustrated in 
Figure \ref{fig:latetime}.
The global features include a dominant accretion flow in the 
equatorial region and mildly supersonic, quasi-periodic outflows
along the two poles.  These outflows are fed by accreting gas that 
``misses the target.''  The oblique shocks near the equator lead to
low-entropy gas advecting in (as material hits the shock at an 
increasingly oblique angle, it does not suffer as large a change
in entropy when traversing the shock).

\begin{figure}[!hbtp]
\plottwo{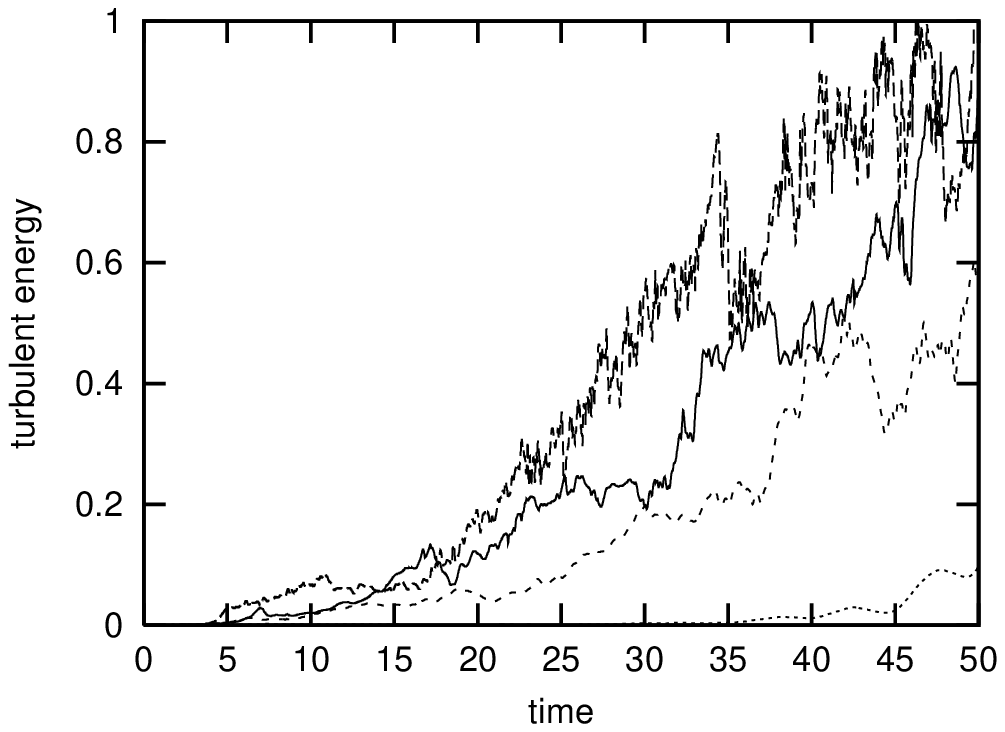}{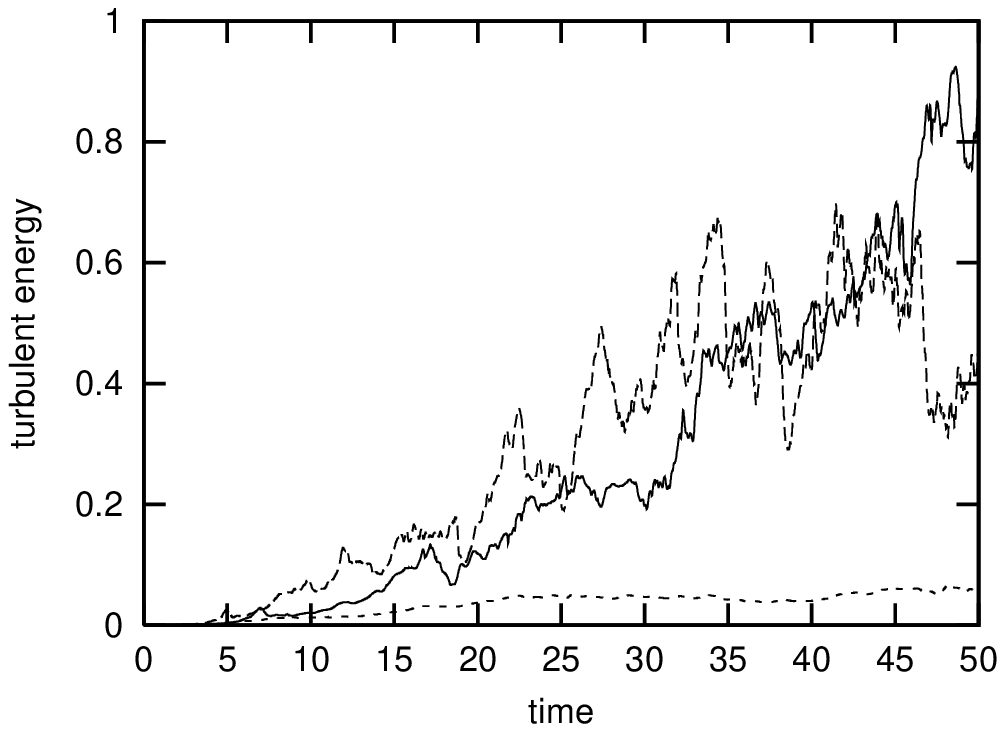}
\caption{
The dependence of the SAS instability on the model parameters is shown 
here by looking at the growth of the interior turbulent energy for
several 
different simulations corresponding to different inner boundary radii 
(left graph: from top to bottom, $R_i$ = 0.05, 0.1, 0.2, and 0.4) and 
different adiabatic indices (right graph: from top to bottom $\gamma$= 
1.4, 1.33, and 1.25, with $R_i$ fixed at 0.1).
}
\label{fig:rad_gam}
\end{figure}

The only two parameters in this idealized SAS model are
the adiabatic index, $\gamma$, and the radius of the inner boundary, $R_i$.
We ran several simulations varying these parameters and found
qualitatively similar results in virtually all the simulations, but
with two clear trends, illustrated in Figure \ref{fig:rad_gam}.  
Simulations with a larger value of $R_i$, and
hence less volume for postshock turbulence to develop, exhibited 
a slower growth in the total turbulent energy. 
\citet{hbc92} pointed out that the scale of convection in a confined
spherical shell scales with the ratio of the inner to the outer boundaries.
By analogy we might then expect the dominant mode of instability
in the accretion shock to scale with $R_i$: the larger the value of
$R_i$, the larger the value of $l$ of the dominant mode.  While the
growth of the instability was slower, all of our models did
eventually evolve into a strong bipolar mode that drove
the shock out to larger radii.  Even the simulation with
$R_i = 0.4$ rapidly grew to a turbulent energy of 0.6 by a time
of 80.  We also found
slower growth of the SAS instability with smaller values of $\gamma$.
Only in the case of a very soft equation of state ($\gamma = 1.25$)
did the shock remain marginally stable for several flow times.  Eventually,
this model became as unstable as the others, 
with a growing bipolar accretion shock.  While all of the SAS models
are unstable, a decrease in the effective adiabatic index and  
spatial extent of the postshock flow does appear to slow
the growth of the instability.  Note that both of these trends might work
against the growth of the 
instability in the evolution of core-collapse supernovae,
where the effective adiabatic index is lower and the postshock volume
is affected by neutrino cooling.

\section{CONCLUSION AND DISCUSSION}

Our primary conclusion is that relatively small-amplitude 
perturbations, whether they originate from inhomogeneities 
in the infalling gas, aspherical pressure waves from the
interior region, or perturbations in the postshock velocity 
field can excite perturbations in a standing accretion shock
that lead to vigorous turbulence and large-amplitude variations in 
the shape and position of the shock front.  

It is important to note that this behavior is not driven by
convection:  the initial conditions in our SAS model are marginally stable to
convection.  Furthermore, during the course of our simulations the
angle-averaged entropy remains relatively constant with radius.  In
Figure \ref{fig:k1a01} we can see the development of convection on
small scales from local entropy gradients in the flow below the
accretion shock. However, there is no global gradient extending
over the entire postshock region, driving convection on that scale.

Our models clearly show an evolution to low-order modes, as first 
seen in the early two-dimensional supernova models of \citet{hbc92}.  
Independent of the initial seed,
our models were eventually dominated by either an $l=1$ or $l=2$ mode
after several flow times; these were the only modes that ever became
highly nonlinear.  
Moreover, once these low-order modes began to
dominate, the average shock radius began to grow, altering the global
behavior from a standing accretion shock to an expanding aspherical
blastwave. 

Our work also suggests that large-scale ($l$ = 1
or 2) modes may dominate the late-time evolution of core-collapse
supernovae, although our models cannot directly address this possibility
because the assumption of supersonic freefall above the shock is no 
longer valid once the shock encounters the subsonic regions
in the stellar core. Nonetheless,
it is interesting to note that when our outflows become
self similar, they achieve an aspect ratio $\sim 2$.
This is within the range suggested by spectropolarimetry
data \citep[e.g.,][]{hkw01}. 
Could it be that the polarization uniformly observed
in core collapse supernovae is a signature of the unstable 
growth of $l=1,2$ modes in the initial explosion? Moreover,
if the SAS instability is responsible for generating the large 
aspect ratios in observed explosions, it provides at least one
example where a ``conventional'' neutrino-driven explosion model
can be constructed that is not in conflict with these observations,
a concern expressed by \citet{wetal01}.

While our results are relatively independent of the initial conditions,
we note that the amplitude and scale of our seed perturbations are
consistent with those expected to be produced by the late stages of
shell burning in presupernova stars \citep{ba98} or in the
core collapse process itself.  Bazan and Arnett followed the evolution
of shell oxygen burning in a presupernova star and found that unstable
nuclear burning can lead to significant density perturbations.  When
this layer of the progenitor star falls through the accretion shock, it
can excite the instabilities found here.  Alternatively, entropy-driven
postshock convection \citep{hbhfc94,bhf95,jm96,mcbbgsu98b,fh00}
could provide the necessary seeds for this shock instability.
Yet another possibility is the growth of lepton-driven convection
\citep{kjm96} or neutron fingers \citep{wm93}
deep in the proto-neutron star.  If these grow to sufficient
amplitude, they would drive aspherical pressure waves out into the accreting
postshock gas, exciting this shock instability.

This instability exists!  We have used time-dependent hydrodynamic
simulations to show that a simple adiabatic model of a standing
accretion shock that is stable in one dimension is unstable in
two dimensions.  The physical mechanism responsible for this 
instability appears to be the feedback between aspherical-shock--induced 
vorticity and the pressure waves generated by this vorticity. This 
feedback was also discovered in the context of accreting black holes 
and dubbed a ``vortical-acoustic'' feedback by
\citet{fog02}.  We note that \citet{hc92} performed a linear stability
analysis of steady, spherical accretion flows with cooling.  We have 
used one of these models to test the effect of different
inner radial boundary conditions on our simulation outcomes.  
For $\gamma=4/3$, our one-dimensional simulations are in
agreement with their results.  In particular, 
we find that the shock is stable for small values of the shock stand-off
distance (when the shock is less than 20 times the radius of the
inner boundary surface) and unstable for larger values.  However,
for $\gamma=4/3$ \citet{hc92} did not extend their analysis to the
nonradial case; therefore, we cannot directly compare their results
with our findings in this case.

Many questions must be answered before we can determine whether this 
vortical-acoustic instability plays an important role in core-collapse 
supernovae.  Will neutrino cooling near the proto-neutron star surface 
dampen the feedback mechanism by damping the pressure waves? Will neutrino 
cooling below the shock drive the flow away from conditions under which 
an instability can develop by reducing the postshock volume?
In the models considered here, radiative losses were neglected, although
mimicked in one case by using a soft ($\gamma=1.25$) equation of state.
In this case the instability was weakened but
still developed into an aspherical flow.

Neutrino cooling and heating will have several important consequences 
for the development of the instability described here: 
(1) As mentioned above, neutrino cooling may affect the generation of 
pressure waves at the base of the density cliff associated with the
surface of the proto-neutron star. 
(2) Neutrino cooling and heating will, in part, determine the postshock 
volume between the density cliff and the supernova shock, which in turn 
will, in part, determine which $l$ modes may grow in the postshock flow. 
(3) Neutrino heating may selectively drive modes other than the fastest 
growing, $l=1,2$, modes.
Moreover, the effects due to neutrino cooling and heating will be 
time dependent. Once an explosion is underway, the cooling region
will be confined to the surface of the proto-neutron star, 
although pressure waves may still be damped, and the heating 
region will increase in volume, allowing lower-$l$ modes to
develop---perhaps the $l=1,2$ modes. 

One can already begin to assess the relevance of the SAS instability 
to the core-collapse supernova problem by comparing our models with 
previous multidimensional core-collapse supernova models. 
In \citet{mcbbgsu98b}, once the shock front is
distorted by the initial neutrino-driven convection, 
the SAS instability does appear 
to affect the postshock flow a few hundred ms after bounce.
The shock front is significantly distorted from the initial 
spherical shape, and this oblique shock feeds strong interior
turbulence.  However, unlike in our SAS models, 
the $l=1$ mode in fact did not come
to dominate the postshock flow. Rather, $l\sim $few modes dominated 
at the shock's peak radius and later were 
``squeezed out'' as the shock receded. In this case, owing to neutrino 
cooling and the lack of sufficient neutrino heating, the separation
between the shock front and the surface of the proto-neutron star
was never sufficient 
for low-$l$ modes of the SAS instability to develop.  To test this
hypothesis we ran a two-dimensional SAS model with an inner boundary
at 0.6, comparable to the effective surface of the proto-neutron star
relative to the shock front at late times in the simulations of 
\citet{mcbbgsu98b}.  While low-$l$ modes did exist, they were
unable to grow significantly in such a confined geometry.  This dependence
on the shock radius with respect to the surface of the proto-neutron star
further emphasizes the need for realistic neutrino heating and 
cooling---i.e., neutrino transport---in multidimensional supernova models.

The simulations of \citet{bhf95} clearly exhibit strong turbulence fed
by an aspherical shock at late times. However, they included only one
quadrant in their explosion simulation, and, therefore, neither an $l$
= 1 or 2 mode was allowed. In the simulations described in
\citet{jm96}, one can again identify a close link between the deformed
accretion shock and strong large-scale vorticies in the postshock flow.
Furthermore, by the end of their simulations, at $\sim 700$ ms, it
appears that the $l=1$ mode is beginning to dominate.  An
interpretation of the SPH models of \citet{hbhfc94} in our context is
complicated by the fact that their simulation is also confined to a
$90^o$ wedge, cutting off the lowest order modes.  Furthermore, their
explosion time scale is relatively short, providing insufficient time
for the long wavelength SAS instability to develop.  This short
explosion timescale is most likely the result of 
their neutrino transport implementation.

It is interesting to note that the instability we identify in this
paper may provide a somewhat revised paradigm for powering core collapse
supernovae. In the models we presented here, explosions were obtained
in the absence of neutrinos, powered by gravitational binding energy.
The SAS instability serves to redistribute the energy of the flow into 
kinetic energy at the expense of internal and gravitational energy.
Thus, both the instability and neutrinos may serve as vehicles to 
convert gravitational binding energy into the kinetic energy of
explosion. If the SAS instability does in fact play a role in 
the explosion mechanism in more realistic models, it would be
interesting to compute, if possible, the contributions to the kinetic energy
in the explosion that result from these two vehicles. Moreover,
it has been pointed out that the energies of explosions powered
by neutrino heating may be limited to $1-3\times 10^{51}$ erg 
\citep{j01}. If the SAS instability sets in and also serves
to power the supernova, we may end up with a more energetic 
explosion, one only partially powered by neutrinos. On another
front, it is interesting to note that the SAS instability 
may help alleviate the precarious balance that is struck during
neutrino shock reheating---between accretion luminosity, mass 
accumulation in the gain region, and ram pressure. As pointed 
out by \citet{j01}, it is necessary to maintain sufficient 
neutrino luminosity and mass in the gain region to power an 
explosion, and this is achieved by maintaining a sufficient 
level of mass accretion through the shock and onto the proto-neutron 
star. On the other hand, this accretion is the source of the ram 
pressure the shock combats in its attempt to propagate outward in 
the stellar core. If neutrino heating is able to sustain a
sufficiently large shock radius for a period of time, the 
SAS instability may act to initiate the explosion, removing 
some of the burden from the neutrinos.

There are other factors that may influence the role of shock instability
in core-collapse supernovae. How will rotation and/or magnetic fields couple 
to this mechanism? For example, will rotation provide sufficient distortion 
of the accretion shock to help drive low-$l$ modes? And will this lead to an 
explosion where otherwise (in the absence of rotation) one would not occur? 
It may be that a combination of near-explosive conditions and rotation will 
be required before the instability we describe can develop. If it does, it 
will certainly contribute to the explosion and should be considered an 
ingredient in the explosion mechanism.

Our results are certainly influenced by the imposed axisymmetry of 
our two-dimensional simulations. For example, our bipolar outflows may
be an artifact of our boundary conditions, although the aspect ratio
we observe when our flows become self similar is certainly an 
independent feature; we could have obtained bipolar outflows with very 
different aspect ratios. Generally speaking, will the SAS instability 
exist in three dimensions?  Will the low-order modes still dominate 
the flow? Will rotation serve the role in three dimensions that the
imposed axisymmetry serves in two? Three-dimensional simulations are 
underway, and we hope to report on them soon.

\acknowledgements 
J.M.B. and A.M. are supported in part by a SciDAC grant from the U.S. DoE 
High Energy and Nuclear Physics Program.
A.M. is supported at the Oak Ridge National Laboratory, managed by
UT-Battelle, LLC, for the U.S. Department of Energy under contract
DE-AC05-00OR22725.
C.D. was supported by an NSF REU program in the Department of Physics
at North Carolina State University.
We thank the North Carolina Supercomputing Center and the Center for
Computational Sciences at ORNL for their generous support of computing 
resources. 
We acknowledge many useful interactions with Raph Hix, Alan Calder,
Jeff Knerr, and Dana Paquin.


\begin{thebibliography}{}        

\bibitem[Bazan \& Arnett(1998)]{ba98}
Bazan, G. \& Arnett, W. D. 1998, ApJ, 496,  316

\bibitem[Burrows, Hayes, \& Fryxell(1995)]{bhf95}
Burrows, A., Hayes, J., \& Fryxell, B. A. 1995, ApJ, 450,  830

\bibitem[Chevalier(1989)]{chev89}
Chevalier, R. A. 1989, ApJ, 346, 847

\bibitem[Colella \& Woodward(1984)]{cw84}
Colella, P., \& Woodward, P. R. 1984, JCP, 54, 174

\bibitem[Foglizzo(2002)]{fog02}
Foglizzo, T. 2002, A\&A, 392, 353

\bibitem[Fryer \& Heger(2000)]{fh00}
Fryer, C. L. \& Heger, A. 2000, ApJ,  541, 1033

\bibitem[Herant, Benz, \& Colgate(1992)]{hbc92}
Herant, M., Benz, W., \& Colgate, S. A. 1992, ApJ, 395,  642

\bibitem[Herant et al.(1994)]{hbhfc94}
Herant, M., Benz, W., Hix, W. R., Fryer, C. L., \& Colgate, S. A. 1994, ApJ, 435,  339

\bibitem[H\"{o}flich, Khokhlov, \& Wang(2001)]{hkw01}
H\"{o}flich, P., Khokhlov, A., \& Wang, L. 2001,
in Proc. of 20th Texas Symposium on Relativistic Astrophysics,
eds. Wheeler, J. C. \& Martel, H. (New York: American Institute 
of Physics)

\bibitem[Houck \& Chevalier(1992)]{hc92}
Houck, J. C. \& Chevalier, R. A. 1992, ApJ, 395, 592

\bibitem[Janka(2001)]{j01}
Janka, H.-T. 2001, A\&A, 368, 527

\bibitem[Janka \& Mueller(1996)]{jm96}
Janka, H.-T. \& M\"{u}ller, E. 1996, A\&A, 306, 167
  
\bibitem[Keil, Janka, and Mueller(1996)]{kjm96}
Keil, W., Janka, H.-T., \& M\"{u}ller 1996, E., ApJL, 473,  L111

\bibitem[Leveque(1998)]{lev98}
LeVeque, R. J. 1998, in 27th Saas-Fee Advanced Course Lecture Notes,
eds. Steiner, O. \& Gautschy, A. (New York:Springer-Verlag)

\bibitem[Mezzacappa et al.(1998)]{mcbbgsu98b}
Mezzacappa, A., Calder, A. C., Bruenn, S. W., Blondin, J. M., Guidry, 
M. W., Strayer, M. R., \& Umar, A. S. 1998, ApJ, 495, 911

\bibitem[Mezzacappa et al.(2001)]{mez01}
Mezzacappa, A., Liebendàrfer, M., Messer, O. E. B., Hix, W. R., Thielemann, F.-K., \& Bruenn, S. W.  2001, PRL,  86, 1935.

\bibitem[Wang et al.(2001)]{wetal01}
Wang, L., Howell, D. A., H\"{o}flich, P., \& Wheeler, J. C. 2002, ApJ, 550, 1030

\bibitem[Wilson \& Mayle(1993)]{wm93}
Wilson, J. R. \& Mayle, R. 1993, Physics Reports 227, 97

\bibitem[Whitham(1974)]{w74}
Whitham, G. B. 1974, Linear and Nonlinear Waves (New York: John Wiley and Sons)

\end{thebibliography}
\end{document}